\documentclass{aa}  

\usepackage{graphicx}
\usepackage{txfonts}
\usepackage{gensymb}
\usepackage{mathtools}
\usepackage{amsmath}
\usepackage{mathrsfs}
\usepackage{natbib}
\usepackage{booktabs}
\usepackage{array}
\usepackage{multirow}
\usepackage{makecell} 
\usepackage{tabularx}
\usepackage{hyperref}
\usepackage{siunitx}
\usepackage{xspace}

\newcommand{\Msunyr}{~M$_\odot$\,yr$^{-1}$\xspace}
   
\newcommand{\ergs}{~erg\,s$^{-1}$}

\begin{document}

   \title{Large-scale emission from $\gamma$-ray binaries: the case of LS~5039}

   \author{J. R. Martinez
          \inst{1,2}
          \corrauth{jmartinez@iar.unlp.edu.ar}
          \and
          V. Bosch-Ramon\inst{3}
          \corrauth{vbosch@fqa.ub.edu}
          }

   \institute{Facultad de Ciencias Exactas, UNLP,
              Calle 47 y 115, CP(1900), La Plata, Buenos Aires, Argentina.\\
         \and
             Instituto Argentino de Radioastronom\'ia (CCT La Plata, CONICET), C.C.5, (1894) Villa Elisa, Buenos Aires, Argentina.
        \and
            Departament de F\'isica Qu\'antica i Astrof\'isica, Institut de Ci\'encies del Cosmos (ICCUB), Universitat de Barcelona (IEEC-UB), Mart\'i i Franqu\`es 1, E-08028 Barcelona, Spain.
             }
   \date{}

% \abstract{}{}{}{}{} 
% 5 {} token are mandatory
 
  \abstract
   {
   Gamma-ray binaries hosting a non-accreting neutron star and a massive star exhibit multi-wavelength emission on different spatial scales. The interaction between the neutron star and stellar winds mixes them and produces an outflow whose interaction with the medium on large scales can inflate a bubble or form a bow shock in systems that are old enough. The well-studied but still mysterious high-energy emitting binary LS 5039 shows extended (1~pc scale) X-rays that might arise from one of these two interaction regimes.
   }
   {
   We explain and predict the large-scale emission from LS~5039.
   }
   {
    We modelled the LS~5039 large-scale emission, including the thermal and non-thermal contributions. For the latter, we assumed particle acceleration at the termination shock of the mixed-wind outflow. We considered five scenarios: a very young and an intermediate-age bubble interacting with the interstellar medium, and a bubble evolving inside a supernova remnant, studied using a one-zone model; and two older bow shocks with a moderate and a high neutron star power, respectively, for which we used a multi-zone approach. We also investigated the particles escaping from these structures and their radiation.
   }
   {
    The large-scale X-rays surrounding LS~5039 are best explained as synchrotron radiation, as previously proposed. In some cases, very extended radio emission might be detectable, while particles escaping the region may be a minor but non-negligible steady contribution to the source gamma rays in the powerful bow-shock scenario. Additionally, escaping protons and nuclei with 0.1--1~PeV might inject up to $\sim 10^{36}$~\ergs\, into the Galactic cosmic rays for optimistic injection luminosities.
   }
  {
  Gamma-ray binaries can be efficient accelerators at large scales and produce non-negligible amounts of broadband emission and 0.1--1~PeV cosmic rays. Our study can help future multi-wavelength observational campaigns to better characterise the large-scale medium interaction, age, and birthplace of LS~5039. 
  }


   \maketitle
   \nolinenumbers
%
%-------------------------------------------------------------------
\section{Introduction}\label{Sec:intro}

Gamma-ray binaries (GBs) consist of a compact object, either a neutron star (NS) or a black hole, and a stellar companion. The multi-wavelength spectrum of GBs is similar to that of X-ray binaries, but peaks in gamma rays \citep{Dubus_2013,Bordas_2024}. About ten GB systems with high-mass stellar companions (which are the most powerful gamma-ray emitters) have been discovered so far: LS~5039 \citep{Paredes_2000,Aharonian_2005_a}, PSR B1259$-$63 \citep{Aharonian_2005_b}, LS~I~+61~303 \citep{Albert_2006}, HESS~J0632$+$057 \citep{Aharonian_2007}, 1FGL~J1018.6$-$5856 \citep{Fermi-LAT_2012}, HESS J1832-093\citep{HESS_2015}, PSR J2032$+$4127 \citep{Lyne_2015}, LMC~P3 \citep{Corbet_2016}, 4FGL~J1405.1$-$6119 \citep{Corbet_2019}, and possibly HESS J1828$-$099 and HESS J1832$-$085 \citep{De_Sarkar_2022,De_Sarkar_2026}. Of these, PSR~B1259$-$63, PSR~J2032$+$4127 and (very likely) LS~I~+61~303 have been confirmed to host a non-accreting NS through the detection of radio pulsations \citep[see][and references therein]{Paredes_2019_a,wen22}.  \cite{Corbet_2016} and \cite{Corbet_2019} suggested that LMC~P3 and 4FGL~J1405.1$-$6119 might also host an NS through the analysis of their light curves and spectra, similarly to HESS~J0632$+$057 and HES~J1832$-$093, which has also been proposed to host an NS \citep[e.g.][and references therein]{Bordas_2024}. The X-ray \citep{Kishishita_2009} and gamma-ray \citep{Aharonian_2006,Abdo_2009,Collmar_Zhang_2014} light curves of LS 5039 along with the absence of accretion-disc features \citep{Martocchia_2005} might also favour the NS scenario. Although this remains unconfirmed, it is also supported by the possible detection of X-ray pulsations (\citealt{Yoneda_2020,Makishima_2023}; see however \citealt{Volkov_2021,Kargaltsev_2023}).

In the (non-accreting) NS scenario, an NS-powered wind interacts with the massive star wind, creating an interaction structure whose geometry, on binary scales, mostly depends on the ratio of the NS-to-star wind momentum rates, $\eta = L_{\rm NS}/\dot{M}_wv_{\rm w}c$, where $L_{\rm NS}$ is the power of the NS wind, and $\dot{M}_{\rm w}$ and $v_{\rm w}$ are the companion mass-loss rate and wind velocity, respectively. Typically, we expect $\eta < 1$, so the interaction structure bends around the NS. The evolution of this structure is complex and has been studied through (semi-)analytical models \citep[e.g.][]{Dubus_2006,B-R_2011,Torres_2012,Bednarek_Sitarek_2013,Molina_2020} and numerical simulations \citep[e.g.][]{Romero_2007,Bogovalov_2008,Bogovalov_2012,B-R_2012,Lamberts_2013,B-R_2015,Barkov_2016,Barkov_2021,Huber_2021_a,Huber_2021_b,Kissmann_2023,Barkov_2024,B-R_2025_b}. The binary orbital motion bends the shocked two-wind structure, which closes and terminates the NS wind in the direction opposite to the star \citep{B-R_2011, B-R_2012}. Several mechanisms lead to the mixing of the two shocked winds. The large velocity gradient between the fast and light shocked NS wind and the slower and heavier shocked stellar wind triggers Kelvin-Helmholtz instabilities on scales $\sim 0.1$--$10\,a$, where $a$ is the binary semi-major axis \citep[e.g.][]{B-R_2012,Lamberts_2013}. Additionally, numerical simulations predict that Richtmyer-Meshkov and Rayleigh-Taylor instabilities occur on the orbit-leading side of the structure \citep{B-R_2015}. All these processes closer to the binary, along with Rayleigh-Taylor instabilities farther out, contribute to the further mixing of the shocked winds at larger scales while they form a disintegrating spiral structure ($\gtrsim 10\,a$) \citep[e.g.][the last work -see also below- emphasising the importance of eccentricity]{B-R_2011,Bednarek_Sitarek_2013,Barkov_2021}. The most recent numerical work of the orbital evolution of the interaction structure and its non-thermal emission on scales close to the binary was developed by \cite{Kissmann_2026}.

Much of the power of GBs is expected to affect their Galactic surroundings, and several GBs present evidence of extended X-ray emission on large scales (LS~I~+61~303, PSR~B1259$-$63, LS~5039, 1FGL~J1018.6$-$5856, and HESS~J0632$+$057; see, e.g. \citealt{par07,pav11,Durant_2011,wil15,kar22}), the origin of which is unclear so far \citep[see, however,][for a study in the case of the very eccentric system PSR~B1259$-$63]{Barkov_2016}. Of this list, LS~5039 is one of the best studied sources, but the nature of its compact object, non-thermal emitter on all scales, and origin (and thus age; see \citealt{B-R_2025_b} for a recent discussion) remain mysterious.  In particular, \cite{Durant_2011} reported an extended X-ray source surrounding LS~5039 with a radius of $\sim$~1'--2', corresponding to a projected physical size of $\sim 0.6$--1.2~pc for a distance of $\approx 2$~kpc \citep{Gaia,Carretero-Castrillo_2026}. Since all the potential energy-dissipation regions (in which particles are accelerated) discussed so far are much closer to the binary than these scales, \cite{B-R_2011} proposed that this emission might originate at the termination shock of an outflow from the binary that consists of mixed stellar and NS winds. The magnetic field, carried by the outflow from binary scales and likely enhanced at the outflow termination site, might cause the accelerated electrons to emit synchrotron radiation from radio to X-rays. Moreover, these relativistic electrons might produce gamma rays via inverse-Compton (IC) scattering of photons from local radiation fields, that is, the cosmic microwave background (CMB), the Galactic infrared (IR) field, and ultraviolet (UV) photons from the LS~5039 star.

Although the uncertainty on the age of the system means that the large-scale interaction regime of LS~5039 is unknown (see Sect.~\ref{Sec:LS_5039} below), the detected extended X-rays can be used to place constraints on some important physical properties of the source. Building upon previous more general work by \citep{B-R_2011}, we developed models of the emission produced by the interaction between the mixed-wind outflow (MWO) and the surrounding medium on large scales and applied them to the case of LS~5039. We considered several possible scenarios to explain the emission reported by \cite{Durant_2011} for different source ages and predicted the thermal and the non-thermal properties in each case. In this paper, we present the (somewhat puzzling) results of this investigation, with each scenario posing a challenge of some sort. The paper is organised as follows: In Sect.~\ref{Sec:large_scale} we discuss the properties of large-scale interaction structures in (pulsar) GBs. In Sect.~\ref{Sec:LS_5039} we address the properties of LS~5039. In Sect.~\ref{Sec:model} we present our model. In Sect.~\ref{Sec:Results} we show our main results, and we summarise our work in Sect.~\ref{Sec:Conclusions}.

%--------------------------------------------------------------

\section{Large-scale interactions in pulsar $\gamma$-ray binaries}\label{Sec:large_scale}

As explained, \cite{B-R_2011} suggested a theoretical framework for the dynamics and non-thermal emission of the large-scale interaction structure in GBs hosting a non-accreting pulsar: The shocked NS and stellar winds are expected to largely mix through instability growth and orbital motion and to turn into a hot, high-entropy turbulent plasma whose mass and energy are determined by the stellar wind and the NS wind, respectively. The strong negative pressure gradient directed away from the binary accelerates this plasma, and it forms the MWO when it eventually becomes highly supersonic. Most of the NS wind power is converted into MWO kinetic power \citep{Barkov_2021}, and at a distance of a few spiral turns, the MWO approaches its maximum velocity, given by
\begin{equation}
    v_{\rm exp} \approx \sqrt{\frac{2\,L_{\rm NS}}{\dot{M}_{\rm eff}}}\,,
    \label{Eq:v_exp}
\end{equation}
with $\dot{M}_{\rm eff}$ being the MWO mass rate. For moderate eccentricities, the resulting flow propagation becomes quasi-spherically symmetric, whereas in high-eccentricity systems, the MWO forms a jet-like structure pointing in the apastron direction \citep{Barkov_2016,Barkov_2021}. For $\eta<1$, the flow in the direction perpendicular to the orbital plane only consists of stellar wind \citep{B-R_2011,Bednarek_Sitarek_2013}, implying that $\dot{M}_{\rm eff}$ is lower than the stellar wind mass-loss rate, $\dot{M}_{\rm w}$. If $\alpha$ is the half-opening angle of the stellar wind--NS wind bow shock, a lower limit to the ratio $\dot{M}_{\rm eff}/\dot{M}_{\rm w}$ can be estimated from \citet{B-R_2011},
\begin{equation}
\frac{\dot{M}_{\rm eff}}{\dot{M}_{\rm w}} \gtrsim\frac{\left(2\,\alpha\right)\,\left(2\,\pi\right)}{4\,\pi} \sim 0.5\,\left(\frac{\eta}{0.02}\right)^{1/3}\,,
\label{Eq:Mdot_eff/Mdot}
\end{equation}
where the expression is valid for small opening angles, such that $\sin{(\alpha)} \sim \alpha$. This estimate is regarded as a lower limit because numerical simulations predict that the stellar wind--NS wind bow shock structure transfers momentum to the stellar wind, causing the shocked structure to expand in the direction perpendicular to the orbital plane \citep{B-R_2015}. We also note that this polar outflow is not powered by the NS wind and is therefore expected to be energetically much less important, so it can be neglected when computing the large-scale non-thermal emission.

The supersonic MWO eventually terminates through a shock produced by the ram pressure balance with the surrounding medium, filling a bubble with shocked MWO that in its turn impinges onto the external medium. Depending on the system history, the external medium can be the interior of the NS supernova remnant (SNR) or the interstellar medium (ISM)\footnote{A narrow time window during which the MWO might interact with a previous stellar wind cannot be discarded because the NS activity can be delayed with respect to the SNR \citep{B-R_2025_b}, but the low stellar wind energetics means that this scenario is a nuanced version of the ISM-interaction scenario.}. When the external medium is the interior of the NS SNR, the evolution of the bubble filled by the MWO is determined by the pressure of the SNR interior because of the very high SNR energy. This pressure is $P_{\rm SNR} = \rho_{\rm ISM}\,v_{\rm SNR}^2$, where $\rho_{\rm ISM} = m_p\,\mu_{\rm ISM}\,n_{\rm ISM}$ is the ISM density, with $m_p$, $\mu_{\rm ISM} \approx 1.3$ and $n_{\rm ISM}$ the proton mass, the mean molecular weight, and the ISM numerical density. In turn, the expansion velocity of the SNR is $v_{\rm SNR} = \frac{{\rm d}R_{\rm SNR}}{{\rm d}t_{\rm NS}}$, with $R_{\rm SNR}$ its radius and $t_{\rm NS}$ the age of the NS. During the adiabatic phase, $R_{\rm SNR} \approx 1.15\,(E_{\rm SNR}/\rho_{\rm ISM})^{1/5}\,t_{\rm NS}^{2/5}$. After the SNR becomes radiative at a radius $R_{\rm rad} \sim 24\,(E_{\rm SNR}/10^{51}~{\rm erg\,s^{-1}})^{1/3}/(n_{\rm ISM}/1~{\rm cm^{-3}})^{-1/3}$~pc, its radius turns into $R_{\rm SNR} \approx (2.8\,E_{\rm SNR}\,R_{\rm rad}^2\,t_{\rm NS}^2/\rho_{\rm ISM})^{1/7}$ \citep{Sedov_1959, Cox_1972, Blinnikov_1982,B-R_2011}. The boundary between the SNR interior (filled with shocked ejecta in this case) and the MWO-filled bubble is determined by the equilibrium between their pressures, and is located at
\begin{equation}
    R_{\rm bSNR} \approx \left({\frac{3\,L_{\rm NS}\,t_{\rm NS}}{10\pi\,\rho_{\rm ISM}\,v_{\rm SNR}^2}}\right)^{1/3}\,.
    \label{Eq:R_bsnr}
\end{equation}
On the other hand, the MWO termination (reverse) shock distance is
\begin{equation}
     R_0 \approx \sqrt{\frac{L_{\rm NS}}{2\pi\,P_{\rm SNR}\,v_{\rm exp}}}\,,
     \label{Eq:R0_SNR}
\end{equation}
and the high sound velocity in the SNR interior impedes the development of a forward shock.

For a GB that is old enough\footnote{or in the unlikely event of a weak SNR due to the direct collapse of a white dwarf into an NS, as speculated in \citealt{B-R_2025_b}.}, the SNR can have already dissipated, or its relic can lie far from the system because of the relative motion of the binary to the SNR. Moreover, the system can have become a strong gamma-ray source some tens of thousands of years after the SNR explosion \citep[i.e. $t_{\rm GB}< t_{\rm NS}$;][and see below]{B-R_2025_b}. In all these cases, the MWO would directly interact with the ISM, which can occur in two different regimes depending on the bubble radius, the binary peculiar velocity, $v_{\rm GB}$, and on $t_{\rm GB}$ \citep{B-R_2011}.

\begin{figure*}
\sidecaption
    \centering
    \begin{minipage}[b]{3cm}
        \centering
        \includegraphics[width=\linewidth]{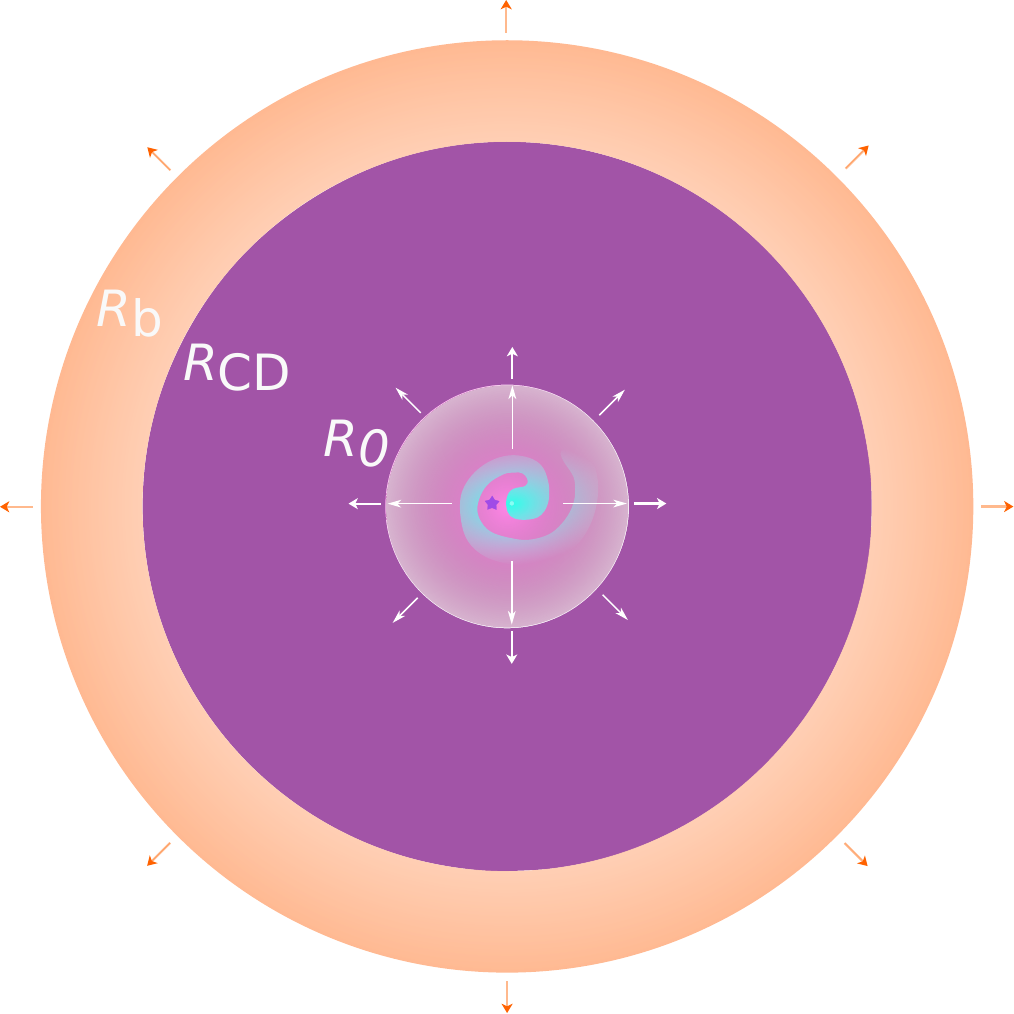}
    \end{minipage}%
    \hspace{0.2cm}
    \begin{minipage}[b]{8.8cm}
        \centering
        \includegraphics[width=\linewidth]{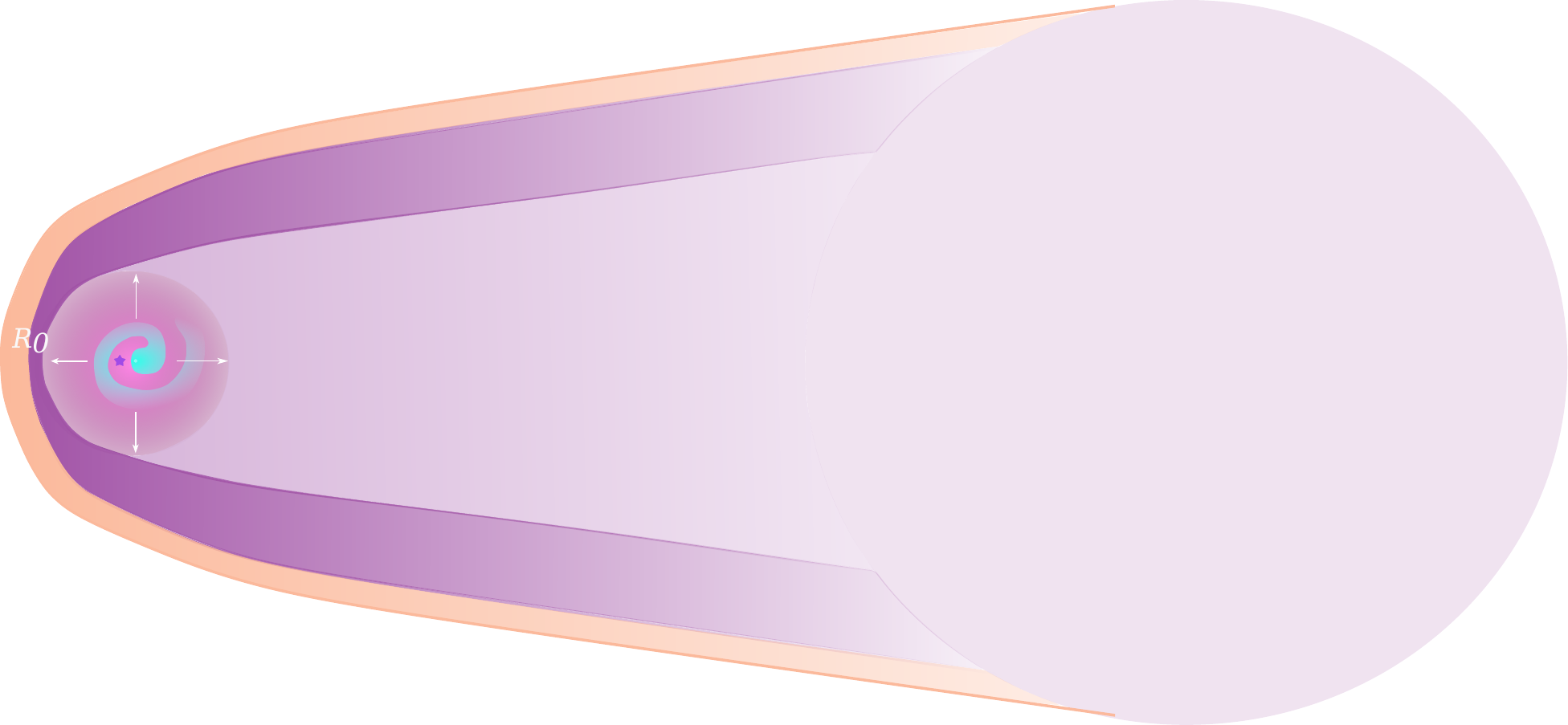}
    \end{minipage}
    \caption{Schematic sketch of the considered scenarios, not to scale.
    {\it Left panel:} Bubble scenario. The MWO is shocked at a distance $R_0$, filling the region up to the contact discontinuity, at $R_{\rm CD}$. When the shocked MWO interacts symmetrically with the ISM, the forward shock sweeps and compresses it.
    {\it Right panel:} Bow-shock scenario. An older GB leaves the bubble due to its motion. The interaction structure is formed by a bow-like and a trailing tail-like structure, the latter ending in a relic bubble filled with shocked and unshocked MWO material.}
    \label{fig:sketch}
\end{figure*}

When $v_{\rm GB}\, t_{\rm GB} < R_{\rm b}$, the interaction structure is roughly symmetric, resembling that of a stellar wind bubble (\citealt{Weaver_1977}; see the left panel of Fig.~\ref{fig:sketch} for a schematic representation). The bubble expands and sweeps the external medium up, which creates a forward shock in the ISM that is adiabatic at early stages ($t_{\rm GB} \lesssim 2\,000$~yr) and later becomes radiative. The bubble radius, including the swept and shocked layer of ISM, is
\begin{equation}
    R_{\rm b} \approx C_{\rm b}\left(\frac{L_{\rm NS}}{\rho_{\rm ISM}}\right)^{1/5}\,t_{\rm GB}^{3/5}\,,
    \label{Eq:R_b}
\end{equation}
with $C_{\rm b}=0.88$ (0.76) for an adiabatic (radiative) forward shock \cite{Weaver_1977}, and its expansion velocity is
\begin{equation}
    v_{\rm b} \approx 0.5\,\left(\frac{L_{\rm NS}}{\rho_{\rm ISM}}\right)^{1/5}\,t_{\rm GB}^{-2/5}\,.
    \label{Eq:v_b}
\end{equation}
In addition, the MWO reverse shock develops at a radius
\begin{equation}
    R_0 \approx 
    \sqrt{\frac{L_{\rm NS}}{2\,\pi\,\rho_{\rm ISM}\,v_{\rm b}^2\,v_{\rm exp}}}\,.
    \label{Eq:R_0_b}
\end{equation}

When $v_{\rm GB}\, t_{\rm GB} \gtrsim R_{\rm b}$, the motion of the GB affects the bubble structure, with the MWO and ISM interaction region becoming bow-shaped (see the right panel of Fig.~\ref{fig:sketch}). The MWO and ISM ram-pressure balance determines the reverse-shock stagnation radius,
\begin{equation}
    R_0 \approx \sqrt{\frac{L_{\rm NS}}{2\pi\,\rho_{\rm ISM}v_{\rm GB}^2\,v_{\rm exp}}}\,.
    \label{Eq:R_0_BS}
\end{equation}
The shocked ISM and MWO flows, separated by a contact discontinuity, form a tail trailing the bow shock. If $t_{\rm GB}$ is barely $\gtrsim R_{\rm b}/v_{\rm GB}$, as considered here because very old NS are unlikely to power luminous GBs, the tail would likely be still bounded by the relic of the original bubble \citep[a similar case is that of an old enough microquasar, as studied in][]{Yoon_2011}.

%%%%%%%%%%%%%%%%%%%%%%%%%%%%%%%%%%%%%%%%%%%%
\section{LS~5039}\label{Sec:LS_5039}

The GB LS~5039 hosts a compact object orbiting an O6.5V((f)) star with a period of $\approx 3.9$~days and an eccentricity of $e \approx 0.35$. The stellar wind velocity and mass-loss rate are $v_{\rm w} \approx 2440$~km\,s$^{-1}$ and $\dot{M}_{\rm w} \sim 10^{-7}$~$\dot{\rm M}_\odot$\,yr$^{-1}$ (e.g. \citealt{Casares_2005}; see also \citealt{B-R_2007}). \cite{Paredes_2000} proposed LS~5039 to be a microquasar hosting an accreting compact object because of the elongated radio structures that are similar to jets. However, several indications suggest that the system might harbour a non-accreting NS. On the one hand, the radio morphology can arise from the region in which the MWO forms \citep[e.g.][]{Dubus_2006,moldon_12b,Molina_2020}. Moreover, \cite{Martocchia_2005}, among others, did not find evidence of accretion features, such as iron line emission or a disc component in the continuum X-ray spectrum. The stability of the X-ray light curve \citep{Kishishita_2009,yon23} and the orbital dependence of gamma rays might also favour an NS scenario \citep{Aharonian_2006,Abdo_2009,Dubus_2010,Collmar_Zhang_2014}. Finally, \cite{Yoneda_2020} reported X-ray pulsations with a period of $\approx 9$~s from LS~5039. Although this result is contested \citep{Volkov_2021,Kargaltsev_2023}, \cite{Yoneda_2020} proposed that this long period suggested that the NS is magnetar-like \citep[see also][]{B-R_2025_b}.

The broadband spectrum of LS~5039 covers radio \citep{Marti_1998,Paredes_2000,Paredes_2002,Marcote_2015}, IR--optical \citep{Clark_2001}, X-rays \citep{Motch_1997,B-R_2007,Takahashi_2009,Hoffmann_2009,Durant_2011}, and gamma rays \citep{Paredes_2000,Aharonian_2005_a,Aharonian_2006,Abdo_2009,Hadasch_2012,Collmar_Zhang_2014,Alfaro_2025}. Although non-thermal modelling showed that synchrotron and IC scattering off stellar UV photons likely cause the whole non-thermal emission, the complex orbital flux and spectral evolution and energetic constraints indicate that the source accelerates particles and emits radiation in multiple regions. For instance, ultrahigh-energy gamma rays (UHE; $\epsilon$ > 100~TeV) might arise from electron--positron pairs accelerated in a magnetised ultrarelativistic outflow within the unshocked NS wind \citep{der12,B-R_2025_a}, whereas the NS wind termination zone and the orbit-induced shock in the NS wind outside of the binary might cause the high-energy (HE) and the very high-energy gamma rays \citep[VHE; e.g.][]{B-R_2011,B-R_2012,Zabalza_2013,Huber_2021_b,Kissmann_2026}. The second acceleration site might also cause part of the source X-rays \citep[e.g.][as some may come from smaller scales]{B-R_2007,Dubus_2010,Szostek_Dubus_2011,B-R_2011}; and radio emission would originate downstream of this site \citep{moldon_12b}. Finally, the large-scale X-rays reported by \cite{Durant_2011} would arise from electrons accelerated at the interaction between the MWO and the medium because no other strong non-thermal sites are expected between MWO formation and termination \citep[see, e.g.][and Sect.~\ref{Sec:large_scale}]{B-R_2011}.

The age and origin of LS~5039 are still unknown, with its runaway nature likely the result of a dynamical ejection and a later supernova explosion \citep{Carretero-Castrillo_2026}. \cite{Ribo_2002} and \cite{Moldon_2012} searched for the birth region of the GB by tracing its past trajectory and reported multiple OB associations and SNRs as candidates, assuming an age of $10^4-10^6$~yr. The origin remained unclear because the uncertainties on the distance and proper motion were large. The distance and velocity of the system were overestimated prior to the advent of the latest {\it Gaia} catalogue \citep{Gaia}. We estimate that the component of the peculiar velocity in the plane of the sky is $v_{\rm GB,t} \sim 90$~km\,s$^{-1}$ taking Galactic rotation into account \footnote{see \url{https://github.com/santimda/intrinsic_proper_motion}.} \citep[as done in][when studying the proper motion of multiple stellar bow shocks]{Martinez_2023}. In addition, the tangential component of the velocity dominates the radial component \citep[][and Carretero-Castrillo, private communication]{Carretero-Castrillo_2023}, so we adopted a peculiar velocity of $v_{\rm GB} = 100$~km\,s$^{-1}$. Nevertheless, despite the refinement of the system peculiar velocity, the age of the binary remains unclear. 

If the non-thermal activity in LS~5039 were fuelled by magnetic dissipation in a very young NS magnetosphere, the source age would seem at odds with the apparent lack of SNR association \citep{Yoneda_2020}. In this regard, \cite{B-R_2025_b} proposed that the NS might be up to a few tens of thousands of years old, corresponding to the duration of a previous weak NS ejector phase, and apparently old enough for the SNR to be already hard to detect. During this NS phase, the binary would have been a faint non-thermal emitter due to a relatively long NS period of a few seconds and the corresponding modest spin-down luminosity. In this scenario, the system would have become a strong non-thermal source after the NS had entered the georotator or propeller regime a few centuries ago, at the moment in which magnetospheric activity would have started to be indirectly induced by the stellar wind. This younger GB-older NS scenario would be associated with a compact bubble interaction regime. Another possibility for an older GB ($t_{\rm GB} \sim 10^4-10^5$~yr) would be plausible if the non-thermal activity were fuelled by the spin-down luminosity of a still powerful pulsar in the ejector phase. Such a scenario would imply a rather short NS period to explain the required high pulsar wind power, so the signal indicating a 9~s X-ray period would be noise or unrelated to the NS rotation. The source in this case might be compatible with the bubble and the bow-shock interaction regimes. In turn, such a bubble might be interacting with a radiative SNR, or (far less likely) directly with the ISM, while a bow shock would have recently abandoned the SNR. 

Although intriguing, the non-detection of an SNR might be explained as follows (a deeper exploration is beyond the scope of this work): when the GB trajectory is traced back using the updated proper motion, the binary crossed a region populated with radio sources 50--100~kyr ago \citep[see][and references therein]{Ribo_2002,Moldon_2012}. This complex environment can mask an old and faint radiative SNR associated with LS~5039. In this case, the lack of a clear SNR counterpart would not be at odds with a standard evolutionary scenario involving a supernova explosion, which is in fact expected according to \cite{Carretero-Castrillo_2026} and remains fully compatible with some of the scenarios discussed here. We add that although disfavoured by indirect evidence (see above), it is still possible that LS~5039 hosts an accreting compact object (either a black hole or an NS) that is $>10^6$~yr old, which hinders tracing its origin. This hypothesis seems less likely at this point and is thus not addressed here.

%-------------------------------------------------------------------
%-------------------------------------------------------------------
%-------------------------------------------------------------------
\section{Model}\label{Sec:model}

Given the large uncertainties on the age of the source, we address the very young and the older NS scenario, which encompass the bubble and the bow-shock interaction regimes. The results obtained were not meant as fits to the data, but rather as illustrative examples showing non-thermal particle and radiation spectra characteristic of each case.

For the bubble scenarios, we considered three alternatives. First, a GB with $t_{\rm GB} \sim 500$~yr (scenario A), compatible with a magnetar-like NS \citep{Yoneda_2020,B-R_2025_b}. This also considers a delayed GB phase from NS formation. Then, a scenario with an intermediate bubble regime representative age of $t_{\rm GB} = t_{\rm NS} \sim 10$~kyr (B). In A and B, the bubble interacts with the ISM. Furthermore, we considered a GB with $t_{\rm GB} \sim 40$~kyr inside an old radiative SNR (C). In this scenario, we set the age of the source such that the SNR already entered its radiative phase, but the termination shock angular scale is compatible with the observational data. We also remark that the radiative nature of the SNR would add to the complexity of the region when trying to explain the SNR non-detection \citep{Ribo_2002,Moldon_2012}. The adopted parameter values, $L_{\rm NS} = 10^{37}$~\ergs, $\rho_{\rm ISM} \approx 2-4 \times 10^{-24}$~g\,cm$^{-3}$, $\dot M_{\rm eff}= 7\times 10^{-8}$~\Msunyr (with $v_{\rm exp}$ being fixed through Eq.~\ref{Eq:v_exp}), and the $t_{\rm GB}$-values determined the size of the bubble.

For the bow-shock scenarios, we considered two different NS powers: $L_{\rm NS} = 10^{36}$~\ergs\, (scenario D) and $L_{\rm NS} = 10^{37}$~\ergs\, (E). In these scenarios, the bow-shock size is independent of the source age. Nevertheless, the system must be old enough to satisfy ($v_{\rm GB}\,t_{\rm GB} \gtrsim R_{\rm b}$) while not being significantly older, since the NS cannot be too old for the considered spin-down luminosities\footnote{As a reference, PSR~B1259--63 has an age of $t_{\rm GB} \approx 330$~kyr and its spin-down luminosity is below $L_{\rm p} \sim 10^{36}$~\ergs \citep[see][and references therein]{Wex_1998,B-R_2011}.}. Thus, we assumed a representative age of $t_{\rm GB} \sim 150$~kyr for scenarios D and E, acknowledging that this might differ by some tens of thousands of years from the real value without affecting our results, except for the emission levels of the relic bubble (see Sect.~\ref{Subsec:model_BS} below). We also note that in D, the NS power would be at the lower limit to explain the non-thermal source luminosity, whereas in E, it might be too high for such an age. In what follows, we explain the models for each interaction regime. The general GB parameters are listed in Table~\ref{table:parameters_general}, and the bubble and bow-shock parameters are presented in Tables~\ref{table:parameters_bubble} and \ref{table:parameters_bow_shock}, respectively.

\begin{table}[t]    
    \centering
    \small
    \caption[]{Model parameters.}
    \begin{tabular}{l l c}
    \hline\hline\noalign{\smallskip}
    \multirow{3}{*}{Parameter}     & \multirow{3}{*}{Symbol}     & \multirow{3}{*}{Value} \\ \\ \hline \\
    GB velocity& $v_{\rm GB}$ & 100 km s$^{-1}$\\
    GB tangential velocity& $v_{\rm GB,t}$ & $95.6\pm 3.2$ km s$^{-1}$\\
    Star mass-loss rate & $\dot{M}_*$ & $10^{-7}$~M$_\odot$\,yr$^{-1}$ \\
    Effective mass rate & $\dot{M}_{\rm eff}$ & $7\times 10^{-8}$~M$_\odot$\,yr$^{-1}$ \\
    \hline\noalign{\smallskip}
    Distance & $d$ & 2.04$\pm$0.06~kpc \\
    \bottomrule
    \end{tabular}
    \tablefoot{GB velocity estimated from \cite{Gaia} and \cite{Carretero-Castrillo_2023}. We calculated the tangential velocity employing the Oort constants given in \cite{Comeron_Pasquali_2007}. The stellar mass-loss rate was based on \cite{Casares_2005}. The fraction of the wind mass loaded into the NS wind ($\dot{M}_{\rm eff}/\dot{M}_\star$) was a free parameter of the model, estimated from the lower limit given in Eq.~(\ref{Eq:Mdot_eff/Mdot}) and constrained semi-quantitatively by the flow evolution in the simulations presented by \cite{B-R_2015}. The distance was calculated from \cite{Gaia}.}
\label{table:parameters_general}
\end{table}

%------------------------------------------------------------------
%-------------------------------------------------------------------

\subsection{Bubble scenarios}\label{Subsec:model_bubble}

In scenarios A and B, the bubble-ISM interaction resembles that of a stellar wind bubble. The shocked-wind region has an inner radius $R_0$ at the MWO reverse-shock location and an outer radius $R_{\rm CD} \approx 0.86\,R_{\rm b}$ \citep{Weaver_1977}, wrapped by a shocked ISM layer. Scenario C is similar, but lacks the shocked-ISM layer. The shocked-flow region is subsonic, and we assumed it to be uniform\footnote{The shocked MWO slows down from a post-shock velocity of $v = 0.25\,v_{\rm exp}$ to a much lower velocity far from the shock; this was neglected.}, allowing us to employ one-zone models for the shocked MWO and the shocked medium. An improved characterisation of the regions employing a multi-zone model would first require the development of 3D hydrodynamical simulations, such as those presented by \cite{B-R_2015} or \cite{Kissmann_2023}, but on larger scales and preferably including the magnetic field, because the actual structure of the shocked MWO region can be very complex.

We assumed the acceleration of non-thermal particles at the MWO reverse shock and computed their emission together with the shocked ISM thermal emission in scenarios A and B. The forward shock in the bubble and the bow-shock scenarios can also accelerate relativistic particles, although we neglected this contribution as the acceleration rate is a factor $(v_{\rm b}/v_{\rm exp})^2 \sim 10^{-9}$ slower than that of the reverse shock. We set the magnetic field $B$ in the shocked MWO region by assuming that the magnetic pressure, $B^2/8\pi$, is a fraction $\eta_B$ of the thermal pressure. This pressure, in turn, is approximately equal to the ram pressure of the MWO immediately before its reverse shock: $P =\rho_{\rm w}\,v_{\rm exp}^2$, with $\rho_{\rm w}\sim\dot{M}_{\rm eff}/(4\,\pi\,R_0^2\, v_{\rm exp})$ (assuming the polar stellar wind region occupies a negligible volume on these scales). We treated $\eta_B$ as a free parameter of the model and constrained its value for each scenario by the X-ray spectrum and by the size of the source (see below).

Particles are accelerated at the MWO reverse shock and injected into the emitting region with a power-law distribution typical of diffusive shock acceleration \citep[$\propto E^{-2}$, e.g.][and compatible with X-ray observations]{Drury_1983}. We adopted the following injection function:
\begin{equation}
    Q\left(E\right) = Q_0 \,E^{-2}\,\exp{\left(-E/E_{\rm max}\right)}\,,
    \label{Eq:Q_inj_Bubble}
\end{equation}
normalised to the non-thermal luminosity injected into protons or electrons,
\begin{equation}
\int_{E_{{\rm min},i}}^{E_{{\rm max},i}} E\,Q_i\left(E\right)\,{\rm d}E = L_{{\rm NT,inj,}i}\,,
\label{Eq:norm_Q}
\end{equation}
where the subscript {\it i} refers to each particle species. The luminosity $L_{{\rm NT,inj},i}$ is a fraction $\eta_i$ of the kinetic power injected by the MWO, constrained by X-ray observations (together with $\eta_{\rm B}$) in the case of electrons. We considered a relation of $\eta_p/\eta_e=10$ as a reference value when estimating the role of protons, although we also explored higher ratios because the proton luminosity is poorly constrained. We set the minimum energy to $E_{{\rm min,}e} = 2\,m_e\,c^2$ for electrons and $E_{{\rm min,}p} = 1$~GeV for protons. Although these minimum energies are uncertain, our results are not expected to change significantly even if $E_{{\rm min,}e}$ were much higher (except perhaps in the radio) because of the modest dependence of $E_{\rm min}$ on the $Q(E)$ normalisation. We determined the maximum energy by equating $\min{\left(t_{\rm GB}, t_{\rm loss}\right)}$, where $t_{\rm loss}$ is the total energy-loss timescale, including radiative and escape losses, to the acceleration timescale. The former is given by
\begin{equation}
    t_{\rm loss}\left(E\right) = \frac{1}{t^{-1}_{\rm cool}+t_{\rm esc}^{-1}}\,,
    \label{Eq:t_loss}
\end{equation}
while the latter is
\begin{equation}
    t_{\rm acc}\left(E\right) = \frac{1}{\eta_{\rm acc}}\frac{E}{c\,q_e\,B}\,,
    \label{Eq:t_acc}
\end{equation}
with $q_e$ being the absolute value of the electron charge, and $\eta_{\rm acc} = (3/8)(v_{\rm exp}/c)^2$ the acceleration efficiency assuming Bohm diffusion \citep{Drury_1983,Protheroe_1999}. We considered Bohm diffusion for the escape losses, with a coefficient $D = E\,c/(3\,q_e\,B)$ and a characteristic timescale of
\begin{equation}
    t_{\rm diff}\left(E\right) = \frac{R_{\rm b}^2}{6\,D}\,.
    \label{Eq:t_diff}
\end{equation}

Electrons radiate through synchrotron emission and IC scattering off stellar UV, CMB, and Galactic IR photons. The synchrotron cooling timescale depends on $B$ and is given by \citep{Pacholczyk_1970}
\begin{equation}
    t_{\rm syn}\left(E\right) = \frac{6\,\pi\,m_e^2\,c^3}{\sigma_{\rm T}}\frac{1}{B^2E}\,,
    \label{Eq:t_syn}
\end{equation}
with $\sigma_{\rm T}$ being the Thomson cross section. In scenario A, where diffusion is similar to synchrotron losses, the X-ray emitting region is (marginally) limited by the bubble size. In scenarios B and C, where radiative losses dominate, electrons can cool well before they diffusively reach the bubble boundary. Therefore, the source will emit up to a radius of $R_{\rm source} \sim R_0 + R_{\rm diff}$, with the diffusion radius being
\begin{equation}
    R_{\rm diff} = \sqrt{\frac{12\,\pi\,\left(m_e c^2\right)^2}{q_e\,\sigma_{\rm T}\,B^3}} \approx 0.55\,\left(B\over30\,\mu{\rm G}\right)^{-3/2}\quad {\rm pc}\,,
    \label{Eq:R_diff}
\end{equation}
determined by equating Eqs.~(\ref{Eq:t_diff}) and (\ref{Eq:t_syn}). This also places an additional constraint on the magnetic field value and encourages a one-zone treatment for these scenarios because the X-ray emitting region is concentrated not far from the termination shock. To calculate the IC losses and emission, we employed the formalism detailed in \cite{Khangulyan_2014}, applying the energy densities given by \cite{Popescu_2017} for the Galactic IR field. Finally, the work exerted on the expanding bubble also causes adiabatic losses in protons and electrons, and the corresponding timescale is $t_{\rm adi} \sim R_{\rm b}/v_{\rm b}$.

We calculated the particle distribution by adapting the time-dependent solution of the transport equation proposed by \cite{Khangulyan_2007},
\begin{equation}
N\left(E,t_{\rm GB}\right) = \frac{1}{|\dot{E}(E)|}
            \int_E^{E_{\rm max}\left(E,t_{\rm GB}\right)}
                Q\left(E'\right) \,{\rm d}E'\,,
\label{Eq:N_1_bubble}
\end{equation}
with $|\dot{E}| = E/t_{\rm loss}$, and with the upper limit $E_{\rm max}$ implicitly defined by the condition
\begin{equation}
\int_E^{E_{\rm max}(E,t_{\rm GB})}
\frac{{\rm d}E''}{|\dot{E}(E'')|}
=
t_{\rm GB}\,.
\label{Eq:Emax}
\end{equation}
In this framework, we treated diffusion escape as a loss term, which is accurate enough for our purposes.

Proton-proton interactions are inefficient to produce significant gamma-ray emission within the bubble. Nevertheless, high-energy protons (and higher-mass nuclei, particularly helium) can diffusively escape from the shocked MWO and interact with the denser ISM (scenarios A and B). The proton losses are negligible within the bubble, and these particles therefore reach the medium with a spectrum similar to $Q_{\rm inj}(E)$ above the minimum energy at which diffusion escape becomes dominant, $E_{\rm diff}$. Considering that protons with energies $E>E_{\rm diff}$ are not trapped inside the bubble, we calculated the distribution of these protons neglecting adiabatic losses,
\begin{equation}
    N_{\rm diff}\left(E\right) \approx Q_{\rm inj}\left(E\right)\,\exp{\left(-E_{\rm diff}/E\right)} \, t_{\rm GB}\,.
    \label{Eq:N_p_diff}
\end{equation}

For the forward shock in scenarios A and B, we calculated the free-free emission employing the formulae given in \cite{Rybicki_1986},
\begin{equation}
    L_{\rm ff}(\epsilon) \propto \frac{n_e\,n_i}{\sqrt{T}\,h} \, e^{-\epsilon/k_{\rm B}T} \,\bar{g}_{\rm ff} \quad {\rm erg\,s^{-1}}\,,
    \label{Eq:L_ff}
\end{equation}
with $n_e \approx n_i \approx 2\,n_{\rm ISM}$ being the numerical density of electrons and ions (taken here to be hydrogen for simplicity), $\bar{g}_{\rm ff}$ is the velocity-averaged Gaunt factor \citep{van_Hoof_2014}, and $k_{\rm B}$ is the Boltzmann constant. We normalised the thermal emission in scenario A taking the volume, ${\mathscr V} = (4\pi/3)\,(R_{\rm b}^3-R_{\rm CD}^3) \approx (4\pi/3)\,(R_{\rm b}^3-0.86\,R_{\rm b}^3) \sim 0.6\,R_{\rm b}^3$, and the thermodynamic properties of the ISM shocked by an adiabatic forward shock into account. This shock is radiative in scenario B, and the material cools within a thin layer. In this case, we normalised the thermal emission by the condition 
\begin{equation}
    \int L_{{\rm ff}\epsilon}\,{\rm d}\epsilon = \frac{\lambda_{\rm ff}\left(T\right)}{\lambda_{\rm cool}\left(T\right)}\,L_{\rm inj,FS}\left(1-\frac{T_{\rm ISM}}{T}\right)\,,
    \label{Eq:Lff_norm_bubble}
\end{equation}
where $L_{\rm inj, FS} = 4\pi\,\rho_{\rm ISM}\,R_{\rm b}^2\,v_{\rm b}^3 \approx 10^{37}$~\ergs\, is the luminosity injected into the shock, $T \approx 1.4\times 10^7\,(v_{\rm b}/10^8~{\rm cm\,s^{-1}})^2~{\rm K}\sim 5 \times 10^5$~K is the immediate post-shock temperature, and $\lambda_{\rm cool}$ and $\lambda_{\rm ff}$ are the total and free-free cooling functions, which can be found in \cite{Myasnikov_1998}.

Finally, we corrected the thermal and non-thermal emission for absorption by the photoelectric effect in the $0.03-10$~keV range. We considered a column density of $N_H = 6.4\times10^{21}$~cm$^{-2}$ \citep{Durant_2011} and the photoelectric cross-sections given in \cite{Morrison_McCammon_1983}. 

%%%%%%%%%%%%%%%%%%%%%%%%%%%%%%%%%%%%%%%%%%%%%%%%%%%%%%%%%

%%%%%%%%%%%%%%%%%%%%%%%%%%%%%%%%%%%%%%%%%%%%%%%%%%%%%%%%%
\begin{table*}[t]    
    \centering
    \small
    \caption[]{Model parameters and brief description of the bubble scenarios.}
    \begin{tabularx}{\textwidth}{p{3.4cm} p{1.8cm} >{\centering\arraybackslash}X >{\centering\arraybackslash}X >{\centering\arraybackslash}X}
    \hline\hline\noalign{\smallskip}
    \multirow{2}{*}{Parameter}     & \multirow{2}{*}{Symbol}     & \multirow{2}{*}{A}  & \multirow{2}{*}{B}     & \multirow{2}{*}{C}\\ 
    {\smallskip}\\ \hline \\
    NS power & $L_{\rm NS}$ [\ergs]                 & $10^{37}$  & $10^{37}$  & $10^{37}$\\
    Expanding wind velocity & $v_{\rm exp}$ [c]                     & 0.07      & 0.07    & 0.07    \\
    \hline\noalign{\smallskip}
    Age of the source & $t_{\rm GB}$ [yr]                     & 500      & 10\,000     & 40\,000    \\
    Reverse shock distance & $R_{\rm RS}$ [$'$]                     & 0.25      & 0.8      & 1.4   \\
    Source radius & $R_{\rm source}$ [$'$]                         & $\approx 1$        & $\approx 1.8$  & $\approx 2$    \\
    Bubble radius & $R_{\rm b}$ [$'$]                               & $\approx 1$        & $\approx 5$     & $\approx 7$         \\
    Bubble expansion velocity & $v_{\rm b}$ [km\,s$^{-1}$]          & $\sim 600$        & $\sim 200$       & $\sim 100$       \\
    \hline\noalign{\smallskip}
    Magnetic field parameter & $\eta_B$                        & 0.01   & 0.1    & 0.5     \\
    Magnetic field  & $B$ [$\mu$G]                              & $\approx 30$   &  $\approx 30$   &  $\approx 40$  \\
    
    $e^-$ injection parameter & $\eta_e$                  & $5\times 10^{-4}$  & $2\times 10^{-4}$    & $2\times 10^{-4}$     \\
    Power injected into $e^-$ & $L_{{\rm inj},e}$ [\ergs]             & $5 \times 10^{33}$   & $2 \times 10^{33}$   & $2 \times 10^{33}$\\
    $p^-$ injection parameter & $\eta_p$                  & $5\times 10^{-3}$  & $2\times 10^{-3}$   & $2\times 10^{-3}$      \\
    Power injected into $p$ & $L_{{\rm inj},p}$ [\ergs]             & $5 \times 10^{34}$   & $2 \times 10^{34}$     & $2 \times 10^{34}$\\
    
    \hline\noalign{\smallskip}
    ISM density     &   $n_{\rm ISM}$ [cm$^{-3}$]           & 1     & 1     & 2 \\
    \hline\noalign{\smallskip}
     & 
        & Pulsar with a $\approx 9$~s period that entered the propeller or the georotator regime $\sim 500$~yr ago. Non-thermal radiation fuelled by magnetic field dissipation in the magnetosphere. Bubble interacting with the ISM.
        &
        Pulsar with a shorter period in the ejector phase. Non-thermal radiation fuelled by the spin-down luminosity. Bubble interacting with the ISM.
        &
        Older NS inside a radiative SNR. Non-thermal radiation fuelled by spin-down luminosity. Bubble surrounded by an old SNR.
    \\
    \bottomrule
    \end{tabularx}
    \tablefoot{The NS power is similar to the upper limit derived by \cite{Zabalza_2011}. The medium density is set according to \cite{Durant_2011}, and also constrained by the large-scale structure size. The magnetic field and injection luminosity are free parameters.}
\label{table:parameters_bubble}
\end{table*}
%-------------------------------------------------------------------
%-------------------------------------------------------------------
\subsection{Bow-shock scenarios}\label{Subsec:model_BS}

To calculate the large-scale emission in bow-shock scenarios D and E, we applied the multi-zone model originally developed by \citet{del_Palacio_2018} and later modified by \citet{Martinez_2022}. The interaction structure was divided into the MWO shocked at its reverse shock and the ISM shocked at the forward shock, both separated by a contact discontinuity. The shocked flows were followed as streamlines propagating away from the bow-shock region, trailing it and forming a tail. We assumed the overall structure to be axisymmetric with respect to the direction of motion, with its shape derived as in \citet{Christie_2016}.

Exploiting the axisymmetry of the model, we considered a longitudinal thin slice of the structure and focused on one half, that is, we started from the stagnation point on the shocked ISM-MWO contact discontinuity. This half slice was approximated as a one-dimensional object or streamline. This streamline was divided into multiple segments, each assigned an angle $\theta$, from the bow-shock apex ($\theta = 0\degree$) to the tail end ($\theta_{\rm max} = 120\degree$, following \citealt{del_Palacio_2018,Martinez_2023}, as almost all of the non-thermal luminosity is injected in this $\theta$-interval). Each streamline in turn consisted of multiple linear emitters divided into cells, with the first cell starting at a different $\theta$ value. Non-thermal particles are injected in the first cell of each linear emitter in the streamline and are subsequently advected to the following cells. We computed the emission from all the cells of all the linear emitters and added their contributions to obtain the emission of the whole streamline. Finally, the emission from the entire structure was obtained by generating the remaining streamlines, reproduced by rotating the streamline around the bow-shock symmetry axis (for more details, see Sect.~2 in \citealt{del_Palacio_2018}).

We considered the same injection function as in Eq.~(\ref{Eq:Q_inj_Bubble}), normalised to the non-thermal luminosity injected at each cell, and determined $E_{\rm max}(\theta)$ as in the bubble scenarios, but including particle advection by the shocked wind with a characteristic timescale given by Eq.~(7) in \cite{del_Palacio_2018} (i.e. the distance from the cell to the system divided by the local shocked flow velocity), further constraining $E_{\rm max}(\theta)$. The steady-state particle distribution at the injection cell was obtained as in that work,
\begin{equation}
N_0\left(E,\theta_l\right) \approx Q\left(E,\theta_l\right)\times \min{\left(t_{\rm cell},t_{\rm loss}\right)}\,,
\label{Eq:N_0}
\end{equation}
where $t_{\rm cell}$ is the advection timescale of the cell. Particles are then advected from one cell to the next, conserving the number of particles within each evolving energy interval as particles lose energy,
\begin{equation}
N\left(\theta_l,E'\right) = N\left(\theta_{l-1},E\right)\,\frac{|\dot{E}\left(E,\theta_l\right)|}{|\dot{E}\left(E',\theta_l\right)|}\,\frac{t_{\rm cell}\left(\theta_l\right)}{t_{\rm cell}\left(\theta_{l-1}\right)}\,,
\label{Eq:N(E)}
\end{equation}
with $E' = E + t_{\rm cell}\,\dot{E}$.

Advected non-thermal particles can eventually reach a relic bubble (see Sect.~\ref{Sec:large_scale}) and produce radiation there, in particular, through synchrotron and IC scattering with the ambient IR and CMB fields. Otherwise, if the bow-shock tail were disconnected from this structure due to tail-disruptive instabilities at late times, non-thermal particles would end up diffusing in the ISM \citep[e.g.][]{Martinez_2023}, which is expected to have similar radiation outcomes. We employed a one-zone model to calculate the relic bubble emission, with an injected electron population roughly following the energy distribution $Q_{\rm relb}\left(E\right) \propto E^{-2}\,\exp{\left(-E/E_{\rm max}\right)}$, normalised by the available power at the last segment of the bow-shock tail, and corresponding maximum energies $E_{\rm max} \sim 10$~TeV (scenario D) and 100~TeV (E) due to synchrotron and diffusive losses, respectively. The steady-state distribution of particles within this relic bubble is
\begin{equation}
    N_{\rm relb}\left(E\right) \approx Q_{\rm relb}\left(E\right) \times \min{\left(t_{\rm GB},t_{\rm loss}\right)}\,,
    \label{Eq:N_ext(E)}
\end{equation}
in which we neglected escape losses due to the expected large bubble size. To compute the resulting emission, we considered $B = 5~\mu{\rm G}$ within the bubble. This value is similar to that in the Galactic disc \citep{Beck_2015}, so the results would still be valid if the particles were injected into the ISM. 

We computed the proton distribution that diffuses into the ISM (see Eq.~\ref{Eq:N_p_diff}) from the tail for the bow-shock D and E scenarios. The radiative forward shock of the bow-shock structure also contributes to the continuum spectrum through free-free emission in these cases. As before, we calculated the emission employing Eq.~(\ref{Eq:L_ff}) and normalised the spectrum with the luminosity injected into each cell, as done in \cite{Martinez_2023}.

%%%%%%%%%%%%%%%%%%%%%%%%%%%%%%%%%%%%%%%%%%%%%%%%%%%%%%%%%%%%%%%%%%%%%%%%%%%%%%%%%%%%%%%%%%%%%%%%

%%%%%%%%%%%%%%%%%%%%%%%%%%%%%%%%%%%%%%%%%%%%%%%%%%%%%%%%%
\begin{table}[t]    
    \centering
    \small
    \caption[]{Model parameters for the bow-shock scenarios. 
    %The NS is in the ejector regime, and its spin-down luminosity fuels the non-thermal activity.
    }
    \begin{tabular}{l l c c}
    \hline\hline\noalign{\smallskip}
    \multirow{2}{*}{Parameter}     & \multirow{2}{*}{Symbol}     & \multirow{2}{*}{D}  & \multirow{2}{*}{E} \\ \\ \hline \\
    NS power & $L_{\rm NS}$ [\ergs]                     & $10^{36}$                     & $10^{37}$   \\
    Expanding wind velocity & $v_{\rm exp}$ [c]                     & 0.07                      & 0.02\\
    \hline\noalign{\smallskip}
    Minimum age of the source & $t_{\rm GB}$ [kyr]                     & $\gtrsim 75$                   & $\gtrsim 100$\\
    Reverse shock distance & $R_{\rm RS}$ [']                     & 0.7                         & 1.3\\
    \hline\noalign{\smallskip}
    Magnetic field parameter & $\eta_B$                        & 0.5                      & 0.05\\
    Apex magnetic field  & B [$\mu$G]                          &  $\sim 35$              &  $\sim 10$\\
    $e^-$ injection parameter & $\eta_e$                     & $0.01$                     & $0.01$\\
    Power injected into $e^-$ & $L_{{\rm inj},e}$ [\ergs]       & $6\times 10^{33}$                & $6\times 10^{34}$\\
    $p$ injection parameter & $\eta_p$                     & $0.1$                     & $0.1$\\
    Power injected into $p$ & $L_{{\rm inj},p}$ [\ergs]       & $6\times 10^{34}$                & $6\times 10^{35}$\\
    \hline\noalign{\smallskip}
    ISM density     &   $n_{\rm ISM}$ [cm$^{-3}$]           & 1     & 1  \\
    \bottomrule
    \end{tabular}
    %\tablefoot{}
\label{table:parameters_bow_shock}
\end{table}

%%%%%%%%%%%%%%%%%%%%%%%%%%%%%%%%%%%%%%%%%%%%%%%%%%%%%%%%%
%%%%%%%%%%%%%%%%%%%%%%%%%%%%%%%%%%%%%%%%%%%%%%%%%%%%%%%%%
%%%%%%%%%%%%%%%%%%%%%%%%%%%%%%%%%%%%%%%%%%%%%%%%%%%%%%%%%
%%%%%%%%%%%%%%%%%%%%%%%%%%%%%%%%%%%%%%%%%%%%%%%%%%%%%%%%%
\section{Results}\label{Sec:Results}

\begin{figure*}
    \centering
    \includegraphics[width=0.99\textwidth]{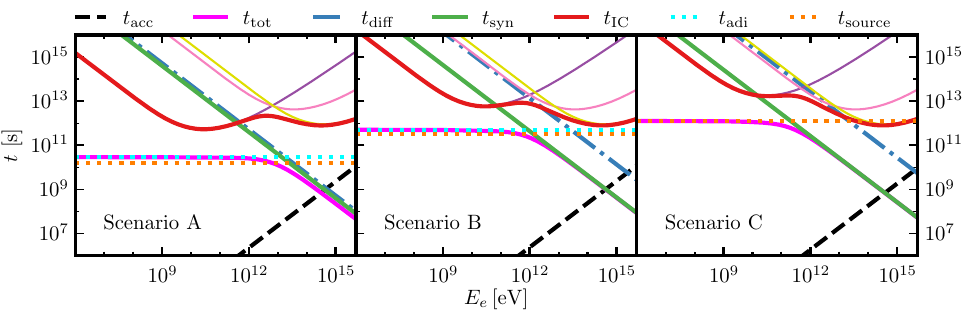}
    \caption{Electron characteristic timescales for bubble scenarios. The solid purple, pink, and yellow lines correspond to IC with the UV, Galactic IR, and CMB photon fields, respectively. The intersection between $t_{\rm acc}$ and $\min{(t_{\rm source}, t_{\rm tot})}$ determines the maximum energy of the particles.}
    \label{fig:tiempos_e_bubble}
\end{figure*}

\begin{figure*}
    \centering
    \includegraphics[width=0.99\textwidth]{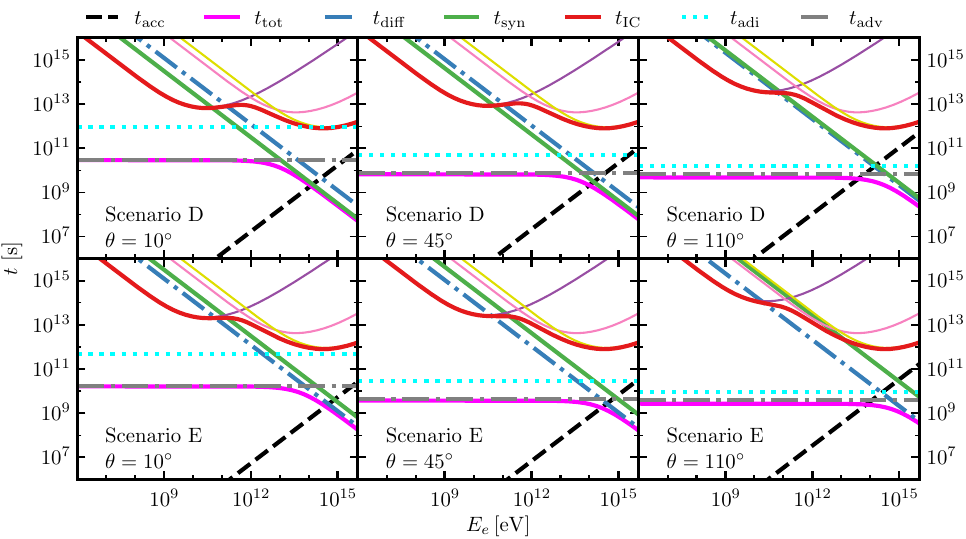}
    \caption{Electron characteristic timescales for the bow-shock scenarios and for different values of the angle $\theta$.}
    \label{fig:tiempos_e_bow_shock}
\end{figure*}

\begin{figure*}
    \centering\includegraphics[width=0.99\textwidth]{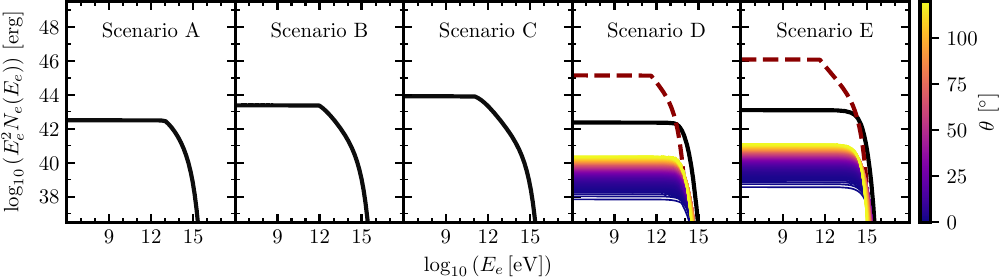}
    \caption{Electron energy distributions. The bubble distributions are strongly constrained by the age of the source and are softened at high energies by synchrotron cooling. We show the distribution as a function of $\theta$ (colour scale) along with the total distribution (black line) in the bow-shock scenarios. The dashed dark red line represents the electron distribution within the relic bubble.}
    \label{fig:diste}
\end{figure*}

\begin{figure*}
    \centering
    \includegraphics[width=0.99\textwidth]{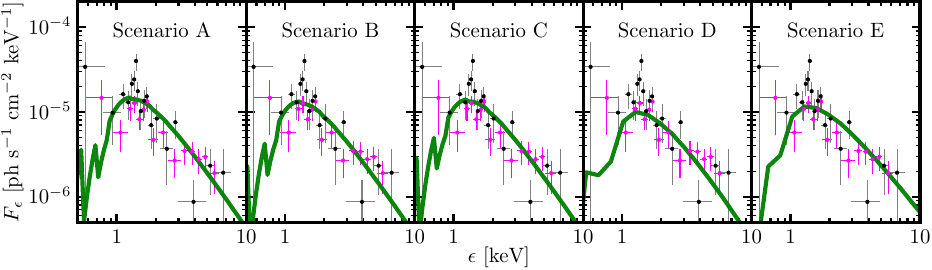}
    \caption{Absorbed X-ray spectrum. We show the \cite{Durant_2011} data up to 1' and 2' in black and magenta, respectively.}
    \label{fig:SED_X-rays}
\end{figure*}

\begin{figure*}
    \centering
    \includegraphics[width=0.99\textwidth]{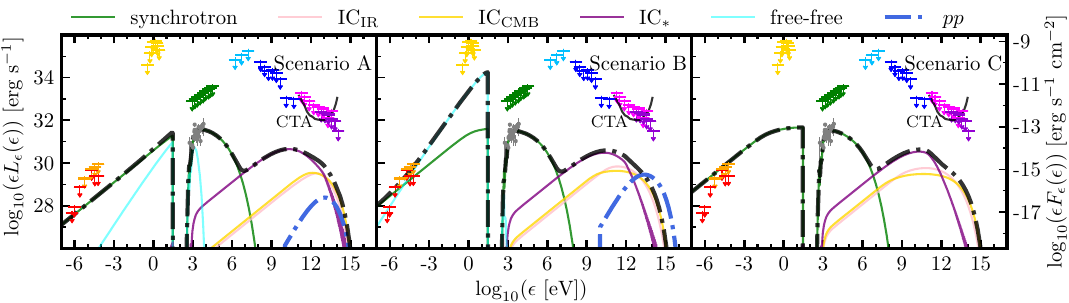}
    \caption{SEDs of the bubble scenarios corrected for photoelectric absorption, along with the \cite{Durant_2011} data in grey. The dash-dotted black line represents the total spectrum emitted within the binary. The dash-dotted blue line represents the proton-proton component from protons that diffuse into the ISM. We considered the data from \cite{Marcote_2015} (red marks) and \cite{Marti_1998} (orange marks) for radio, \cite{Clark_2001} (yellow marks) for optical, \cite{Takahashi_2009} (green marks) for X-rays, \cite{Collmar_Zhang_2014} (light blue marks) for $\sim 10$~MeV, \cite{Hadasch_2012} (blue marks) for HE, \cite{Aharonian_2006}(magenta marks) for VHE, and \cite{Alfaro_2025} (violet marks) for UHE as upper limits. These are also the minimum luminosities across the binary orbit at each spectral band. We took the CTA 50~h observation time sensitivity curve from \cite{CTA}.}
    \label{fig:SED_bubble}
\end{figure*}

\begin{figure*}
    \centering
    \sidecaption
    \includegraphics[width=12cm]{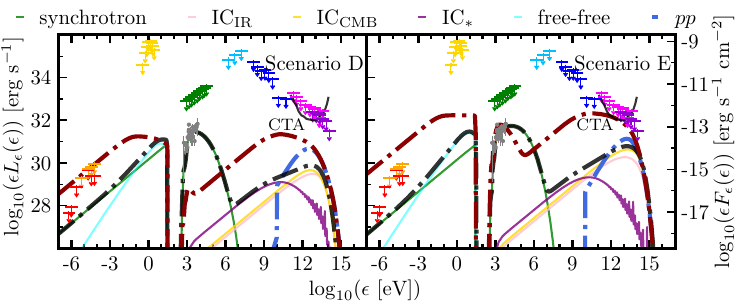}
    \caption{SEDs of the bow-shock scenarios corrected for photoelectric absorption. The dash-dotted black, dark red, and blue lines represent the total SED from the bow shock, the relic bubble, and from protons diffusing through the ISM, respectively. The solid lines represent the radiation through each mechanism in the bow shock.  We considered the same data as in Fig.~\ref{fig:SED_bubble}.}
    \label{fig:SED_bow_shock}
\end{figure*}

The characteristic electron timescales for the bubble (A, B, and C) and bow-shock (D and E) scenarios are shown in Figs. \ref{fig:tiempos_e_bubble} and \ref{fig:tiempos_e_bow_shock}, respectively, and all the energy distributions in Fig.~ \ref{fig:diste} (from left to right). Figure~\ref{fig:SED_X-rays} presents the X-ray data from \cite{Durant_2011}, together with the synchrotron emission (corrected for photoelectric absorption) predicted for all these scenarios. They agree reasonably well. The multi-wavelength spectral energy distributions (SEDs), together with the extended X-ray data and the observed emission at other bands (presumably from smaller scales, so giving upper limits for the large-scale emission of the system) are presented in Figs.~\ref{fig:SED_bubble} and \ref{fig:SED_bow_shock}.

In the young GB case (scenario A), the electron energy distribution is not cooled because the source is young, up to $E \sim 20$~TeV. The most energetic electrons, on the other hand, barely reach the bubble boundary via diffusion because they cool via synchrotron. This cooling softens the energy distribution of these electrons such that their large-scale synchrotron X-rays match the observations \citep{Durant_2011}. The value of $\eta_B = 0.01$ adopted as a representative value in this scenario is relatively low to match the observed source size, although we note that an even lower magnetic field with a higher luminosity injected into electrons is also possible. In that case, the most energetic electrons would diffusively escape from the bubble, suitably softening the X-ray-emitting electron distribution. 

Scenarios B and C allow for a larger and older bubble. This implies longer diffusion timescales, with electrons trapped inside the bubble, and synchrotron losses determining the size of the X-ray emitter. Electrons with $E \gtrsim 1$~TeV (200~GeV) are synchrotron-cooled and reach a radius of approximately $2'$ for a magnetic field of $B = 30\mu{\rm G}$ ($40\mu{\rm G}$) in scenario B (C). This roughly agrees with observations. 

In scenario D, only electrons with energies above $E\sim 10$~TeV cool via synchrotron, and this emission alone causes the relatively soft extended X-rays in this case, whereas beyond $\theta \approx 45\degree$, advection starts to dominate at all energies. In scenario E, the smaller magnetic field results in a diffusion-dominated regime at high energies near the head of the bow shock, with diffusion helping to reproduce the soft extended X-rays, whereas advection again dominates for $\theta \gtrsim 45\degree$ at all energies. Once again, we remark that a different magnetic field of $B \approx 30~\mu{\rm G}$ with a lower $\eta_e$ is also possible, hindering diffusion into the ISM. Electrons advected through the bow-shock tail accumulate in the relic bubble. Their energy distribution is easily characterised by the distribution at the tail up to hundreds of GeV, while it softens through synchrotron cooling above these energies.

\begin{figure*}
    \centering\includegraphics[width=0.99\textwidth]{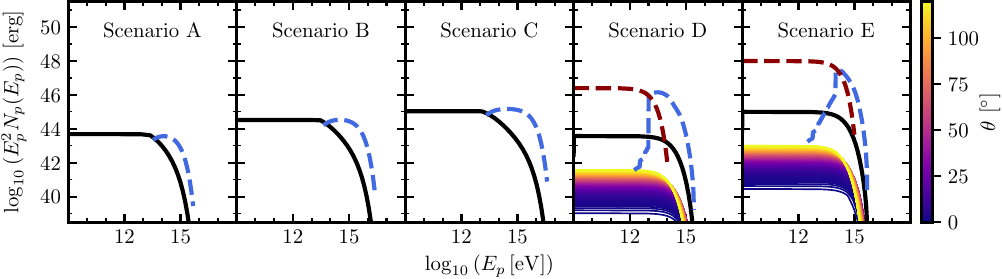}
    \caption{Proton energy distributions. The solid black line refers to the total distribution within the shocked MWO. The dashed dark red line represents the proton distribution within the relic bubble. The dashed blue line refers to the total distribution of protons that diffuse into the medium.}
    \label{fig:distp}
\end{figure*}

Figure~\ref{fig:distp} shows the energy distributions of the relativistic protons in the shocked MWO or escaped into the ISM/SNR for all the scenarios and of those inside the relic bubble in the bow-shock scenarios. Protons (and more easily, helium nuclei, not shown) can reach PeV energies under the assumed conditions, making LS~5039 a PeVatron. This is particularly true in scenarios B and C, where the absence of advection losses and the larger bubble size imply a longer residence time in the acceleration region. We note, however, that in C, these protons are still confined by an SNR in its radiative phase, and some adiabatic losses are possible before they eventually escape into the ISM. The normalisations of the proton energy distributions in Fig.~\ref{fig:distp} were determined by the choice $\eta_p=10\,\eta_e$, with $\eta_e\sim 10^{-4}$--$10^{-3}$ in these scenarios (see Table~\ref{table:parameters_bubble}). However, $\eta_p$ can be much higher. For instance, the total non-thermal energy in protons above 100~TeV can be as high as $\sim 2 \times 10^{36}$~\ergs for an optimistic value of $\eta_p=0.5$. 

In the bow-shock scenarios, protons with the highest energies diffuse away near the apex while those with $E > 10$~TeV ($E > 100$~TeV) escape from the trailing tail in scenario D (E). In scenario E, the injected power above 100~TeV is $L_{\rm inj} \sim 3 \times 10 ^{34}$~\ergs\, for $\eta_p = 0.1$, or $\sim 2 \times 10^{35}$~\ergs\, for $\eta_p = 0.5$. The $L_{\rm inj}(E >100~{\rm TeV})$ value in scenario B is ten times higher than in E for three reasons. First, protons can reach higher energies in the bubble. Second, only $\sim 60$–$70\%$ of the MWO is strongly shocked in the bow shock, and we did not consider particle acceleration in the MWO launched opposite to the proper motion direction. Third, the shock becomes increasingly oblique at large $\theta$, further reducing the injected luminosity. Despite all this, scenario E shows the highest proton-proton gamma-ray luminosity from cosmic rays diffusing into the ISM because the longer $t_{\rm GB}$ implies a larger number of protons in the surroundings (see the proton distributions and SEDs below). An extreme case with $\eta_p=0.5$ would yield a luminosity close to the luminosity found by HAWC at superior conjunction (when the lowest emission at binary scales takes place) \citealt{Alfaro_2025}.

As expected, extended X-ray emission is easily explained by synchrotron radiation and far more difficult to explain by IC scattering: it would require a rather steep electron energy distribution with a very high energetics because IC scattering is inefficient at the relevant electron energies. The predicted IC gamma-ray luminosities are lower than that of the binary, although we predict a potential marginal detection of the emission from the relic bubble in scenario E with the Cherenkov Telescope Array.

In scenarios B and C, the total radio fluxes at hundreds of megahertz are above the total upper limits reported by \cite{Marcote_2015}. However, these upper limits correspond to a much smaller scale. At the synthesised beam of 18" of the Giant Metrewave Radio Telescope, the upper limits correspond to a surface brightness of $S_{\rm 154\,MHz} < 60$~mJy \,arcmin$^{-2}$, consistent with our predicted values of $S_{\rm 154\,MHz} \approx 3$~mJy\,arcmin$^{-2}$ and $S_{\rm 154\,MHz} \approx 20$~mJy\,arcmin$^{-2}$ for scenarios B and C, respectively. The free-free emission from the radiative forward shock is highest in scenario B, with its thermal spectrum peaking in the UV. Nevertheless, this emission would still be, like the star radiation, strongly absorbed in the near-UV, and thus, it would be undetectable.

A single-dish radio telescope is required for arcminute-scale observations (bubble or bow shock) or for 10~arcminute scales (SNR or relic bubble). The Parkes Observatory\footnote{See instrument calculator at \url{https://www.parkes.atnf.csiro.au/cgi-bin/utilities/pks_sens.cgi}.} is well suited for such observations given its southern location and large beam size of $\sim 5'$ at 5~GHz. In scenario D, the relic bubble would be undetectable, while in scenario E, its radio emission lies above the instrument sensitivity. However, it would likely remain undetectable because it is located in a region populated by brighter radio sources \citep[see][and references therein]{Ribo_2002,moldon_12b}, similarly to the putative SNR discussed in Sect.~\ref{Sec:LS_5039}. On the other hand, scenarios B and C are most prominent on arcminute scales. However, the source was not detected in previous surveys \citep{Tasker_1994}, and we estimate that even with deeper observations (total integration times of $\sim 1$~h), the expected detection significance would remain marginal ($\sim$2--3~$\sigma$).

For the X-ray morphology, we note that softer and fainter emission is detected at low significance on scales between $1'$ and $2'$ \citep{Durant_2011}. Nevertheless, given the approximate nature of the adopted models and the lower statistics beyond $1'$,
our results seem to reproduce the extended X-ray observations reasonably well at a semi-quantitative level. The morphology reported by \cite{Durant_2011} might favour a bubble scenario because the emission profile does not peak in the direction of motion of the system. Nonetheless, as with the source extension, this might arise from insufficient statistics or from a non-trivial geometry of the MWO itself, expected to be anisotropic due to the complexity of the colliding-wind region and the (moderate) system eccentricity \citep{B-R_2015,Barkov_2021}.

%%%%%%%%%%%%%%%%%%%%%%%%%%%%%%%%%%%%%%%%%%%%%%%%%%%%%%%%%
\subsection{Future observational tests}
\label{Subsec:future_observations}

The age and origin of LS~5039 remain uncertain. Moreover, the large-scale X-rays can be explained by scenarios with ages spanning from centuries to several tens of thousands of years. However, a deeper study of the large-scale interaction might shed light on the discussion.

The discovery of a bow shock and an elongated non-thermal structure, potentially ending in a gamma-ray emitting relic bubble offset from the binary, would suggest a bow shock regime. This would also either imply an old source that abandoned its SNR or an old source within a diluted SNR. On the other hand, if it were confirmed that the whole interaction structure is bubble-like and has a physical size of $\sim 1'$ with deeper radio and X-ray observations, a very young magnetar-like pulsar would be favoured. In contrast, a larger radio bubble with a size of $\sim 5-10'$ would be compatible with an older NS. The diffuse large-scale emission must be separated from the compact emission associated with the binary to distinguish between these scenarios. The improved resolving power of the next generation of radio facilities might make this possible and might also unveil the presence of an SNR. This would place valuable constraints on the nature and evolutionary stage of LS~5039.

%%%%%%%%%%%%%%%%%%%%%%%%%%%%%%%%%%%%%%%%%%%%%%%%%%%%%%%%%
%%%%%%%%%%%%%%%%%%%%%%%%%%%%%%%%%%%%%%%%%%%%%%%%%%%%%%%%%
%%%%%%%%%%%%%%%%%%%%%%%%%%%%%%%%%%%%%%%%%%%%%%%%%%%%%%%%%
\section{Summary and conclusions}\label{Sec:Conclusions}

We modelled the arcminute-size X-ray emission found around LS~5039. These extended X-rays can arise from the large-scale interaction between a supersonic non-relativistic outflow, formed close to the binary via mixing of NS and stellar winds, with the surrounding medium. The interaction regime mostly depends on the age of the source: for a young source, the interaction is quasi-spherical, whereas for an older source, it would be bow-shaped due to the peculiar velocity of the system. Because the source history is poorly constrained, with the age of the GB phase spanning from a few hundred years up to $\sim 10^5$~yr, we considered both regimes.

The extended X-rays seem to be produced by multi-TeV electrons accelerated at the reverse shock of the MWO with a modest conversion of kinetic into non-thermal power, and emitting via synchrotron radiation in a magnetic field of 10--40~${\rm \mu G}$. Although the X-ray morphology might seem to favour the bubble scenarios over the bow-shock scenarios, the observational properties can be roughly reproduced in both cases. We remark that deeper radio and X- and gamma-ray observations would be useful to determine the large-scale morphology and thus the evolutionary state of the source. A compact ($\sim 1$--$2'$) size of the whole MWO-medium interaction structure would favour a magnetar-like NS powered by magnetic activity, whereas the detection of a larger bubble or a bow-shock morphology ending in a gamma-ray-emitting relic bubble would indicate a system powered by the pulsar spin-down luminosity.

We emphasise that all the considered scenarios have some exotic element that can be hard to accommodate in an approximating description of the source behaviour. All the scenarios satisfy the constraints imposed by the overall SED of the source and explain the extended X-rays, however: scenario A requires a GB phase that is delayed by several tens of thousands of years because no SNR is observed, whereas scenario B requires an SNR that dissipates on an even shorter timescale (or perhaps, even less likely, that does not occur at all). Scenario C might be less problematic in this respect, particularly if the non-detection of the SNR is due to a complex radio environment. Scenario D requires a very high non-thermal efficiency to explain the total non-thermal SED of the source with such a modest MWO luminosity, which is on the other hand more fitting for an older age (although this case might be the least demanding, possibly together with C). Finally, scenario E requires a very powerful ejector regime in a pulsar that is already $\sim 100$~kyr old, which might be unrealistic. This apparently paradoxical situation within the a priori preferred non-accreting NS framework, in which none of the explored options might seem particularly adequate, warrants further observational studies of the large-scale structure generated by the interaction between LS~5039 and its surrounding medium. Such observational studies, together with the modelling of their results, would be a complementary avenue to clarify the nature of this mysterious source.

%--------------------------------------------------------------------

%--------------------------------------------------------------------

\begin{acknowledgements}
We want to thank the referee for valuable and constructive comments that helped to improve the clarity of the manuscript.
We thank Dr. Mar Carretero-Castrillo for providing preliminary results that helped constrain the 3D velocity of LS~5039, and Dr. Dmitry Khangulyan for useful suggestions and helpful discussions. 
J.R.M. acknowledge support by PIP 2021-0554 (CONICET).
VB-R acknowledges financial support from the State Agency for Research of the Spanish Ministry of Science and Innovation under grants PID2022-136828NB-C41 and CEX2024-001451-M funded by MICIU/AEI/10.13039/501100011033/ERDF/EU. V.B-R. is a Correspondent Researcher of CONICET, Argentina, at the IAR.
      
\end{acknowledgements}

\bibliographystyle{aa} 
\bibliography{bibliography}

%-------------------------------------------------------------------%-------------------------------------------------------------------

%-------------------------------------------------------------------%-------------------------------------------------------------------

\end{document}